# Virtual vs. Reality: External Validation of COVID-19 Classifiers using XCAT Phantoms for Chest Computed Tomography

Fakrul Islam Tushar[1], Ehsan Abadi[1], Saman Sotoudeh-Paima[1], Rafael B. Fricks[1,2], Maciej A. Mazurowski[1], W. Paul Segars[1], Ehsan Samei[1], Joseph Y. Lo[1]

[1]Center for Virtual Imaging Trials,
Carl E. Ravin Advanced Imaging Laboratories,
Dept. of Radiology, Duke University School of Medicine
Dept. of Electrical & Computer Engineering, Pratt School of Engineering, Duke University

[2]National Artificial Intelligence Institute, U.S. Dept. of Veterans Affairs.

**Abstract**

Research studies of artificial intelligence models in medical imaging have been hampered by poor generalization. This problem has been especially concerning over the last year with numerous applications of deep learning for COVID-19 diagnosis. Virtual imaging trials (VITs) could provide a solution for objective evaluation of these models. In this work utilizing the VITs, we created the CVIT-COVID dataset including 180 virtually imaged computed tomography (CT) images from simulated COVID-19 and normal phantom models under different COVID-19 morphology and imaging properties. We evaluated the performance of an open-source, deep-learning model from the University of Waterloo trained with multi-institutional data and an in-house model trained with the open clinical dataset called MosMed. We further validated the model's performance against open clinical data of 305 CT images to understand virtual vs. real clinical data performance. The open-source model was published with nearly perfect performance on the original Waterloo dataset but showed a consistent performance drop in external testing on another clinical dataset (AUC=0.77) and our simulated CVIT-COVID dataset (AUC=0.55). The in-house model achieved an AUC of 0.87 while testing on the internal test set (MosMed test set). However, performance dropped to an AUC of 0.65 and 0.69 when evaluated on clinical and our simulated CVIT-COVID dataset. The VIT framework offered control over imaging conditions, allowing us to show there was no change in performance as CT exposure was changed from 28.5 to 57 mAs. The VIT framework also provided voxel-level ground truth, revealing that performance of in-house model was much higher at AUC=0.87 for diffuse COVID-19 infection size >2.65% lung volume versus AUC=0.52 for focal disease with <2.65% volume. The virtual imaging framework enabled these uniquely rigorous analyses of model performance, which would be impracticable with real patients.

## Introduction:

Application of deep learning-based algorithms on the medical imaging-based diagnosis of COVID-19 has resulted in innumerable studies often with nearly perfect performance [1-8]. Despite some studies using moderately large datasets of hundreds to over a thousand cases, most still suffer from serious overtraining common to artificial intelligence (AI) in medical imaging. Even when trained carefully, model performance can vary depending on data population, image quality, image acquisition protocols, dose differences, and disease appearance [9]. These concerns greatly reduce the generalizability of most AI research studies, making them impractical for clinical translation [10, 11].

Virtual imaging trial (VIT) is a process of simulating imaging evaluations with varying factors such as computational human phantoms, imaging scanner systems, and virtual readers [12, 13]. The above AI model belongs in the virtual reader category. VIT can offer practical solutions to evaluate these AI models while rigorously controlling for the virtual patient and scanner conditions, which is nearly impossible with actual patients and scanners. In a prior study, we first described phantoms with COVID-19 that could be imaged using scanner-specific computed tomography (CT) and chest x-ray simulators [14]. To evaluate that simulation framework, we assessed an open-source C0VID-19 diagnostic model and an in-house model, both of which underwent external testing on our simulated CT scans as well as another clinical dataset. The in-house model was intended to provide 3D classification but using an intentionally lightweight model to minimize overtraining. To the best of our knowledge, this was the first study that explored the usability of VIT in evaluating AI models' performance on COVID-19 classification in CT.

**Methods:**

Fig.1 showed the overall workflow of this study. This study from the University of Waterloo used an open-source deep learning model [2] trained and validated on a dataset called COVIDx-CT [2]. We chose this study because it was the largest study where the data, code, and model were all publicly available. We have also developed an in-house 3D COVID-19 classifier utilizing the open-source MosMed dataset [15]. Utilizing both trained models, we evaluated the performance on simulated COVID-19 and normal CT images, which were developed using the VIT process developed and described in detail [14]. Furthermore, we also performed external testing on COVID-CT-MD, which is another public clinical COVID-19 dataset [16].

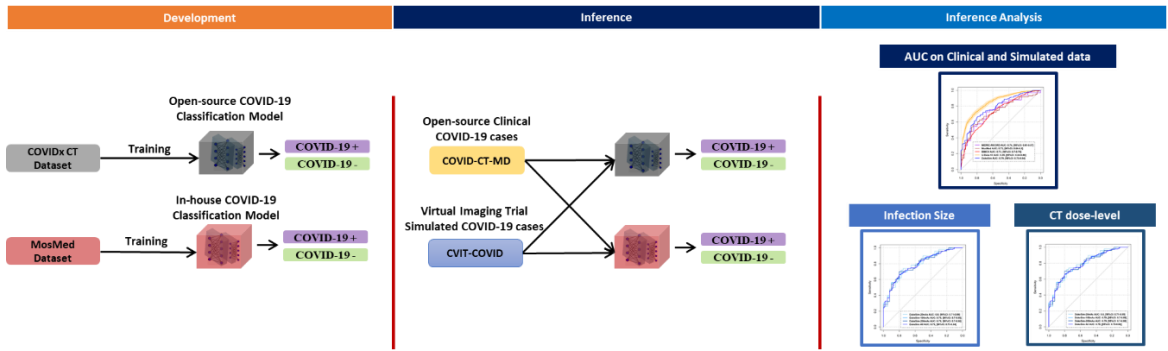

**Figure 1.** Overall workflow. Open-source and in-house deep learning models were externally validated on simulated and clinical COVID-19 data based on patient-level AUC.

**Dataset and Preprocessing:**

In our VITs framework, the 4D extended cardiac-torso (XCAT) phantoms were implanted with simulated disease [14]. Using 50 CT images from patients having confirmed COVID-19, the morphological characteristics of the disease were segmented and simulated into the XCAT models, which then underwent simulated CT scanning. Table 1 details the number of volumes and the different dose levels of the virtual images of CT images which we denoted as Duke center for virtual imaging CT (**CVIT-COVID)** dataset. This study also utilized a publicly available patient COVID-19 dataset called COVID-CT-MD, which included 169 COVID-19 positive and 76 normal and 60 pneumonia CT scans, or a total of 305 CT images [16]. Fig. 2 shows CT slices from simulated XCAT and the clinical COVID-CT-MD datasets. Furthermore, to develop the in-house COVID-19 classifier, we have utilized 1110 CT volumes of MosMed dataset, and were randomly divided by patient into subsets to train (60%), validate (20%), and test (20%) the model [15]. Table 2 shows the numbers of COVID-19 positive and negative instances of COVID-CT-MD [16] and MosMed [15] datasets. COVID-CT-MD is a public COVID-19 dataset introduced by Afshar *et al.*, consisting of not only COVID-19 cases, but also healthy and participants infected by Community Acquired Pneumonia (CAP) [16]. MosMed is an open-source COVID-19 CT dataset consist of COVID-19 positive/negative findings acquired from municipal hospitals in Moscow, Russia by Morozov *et al.* [15].

Similar preprocessing was performed to individual slice images described by the Waterloo study [2]. The CT slices were cropped to the body region, and intensities values were converted in the range between 0-1. For the in-house model, CT volumes were resampled to a voxel size of $5\ mm\ \times 2\ mm\ \times 2\ mm$ by B-spline interpolations and clipped to intensity range in Hounsfield units $(-1000, 800)$. 3D patches of size $96\ \times 160\ \times 160\ (Z \times H\ \times W)$ used for training in-house model and patch sizes were based on organ size plus a margin to allow for patient variability and to include most or all of the chests.

**Table 1.** Duke center for virtual imaging CT (**CVIT-COVID**) dataset. Simulated COVID-19 and normal CT images with different CT dose levels.

|  | Number of volumes | |
| --- | --- | --- |
| Dose level | COVID-19 positive | Covid-19 negative (Normal) |
| 28.5 mAs | 50 | 40 |
| 57 mAs | 50 | 40 |
| **Total** | **100** | **80** |

**Table 2.** Data distribution for COVID-CT-MD described in Afshar *et al.* [16] and MosMed by Morozov *et al.* [15] clinical datasets.

| Dataset | COVID-19 positive | Covid-19 negative |
| --- | --- | --- |
| COVID-CT-MD [16] | 169 | 136 |
| MosMed [15] | 856 | 254 |

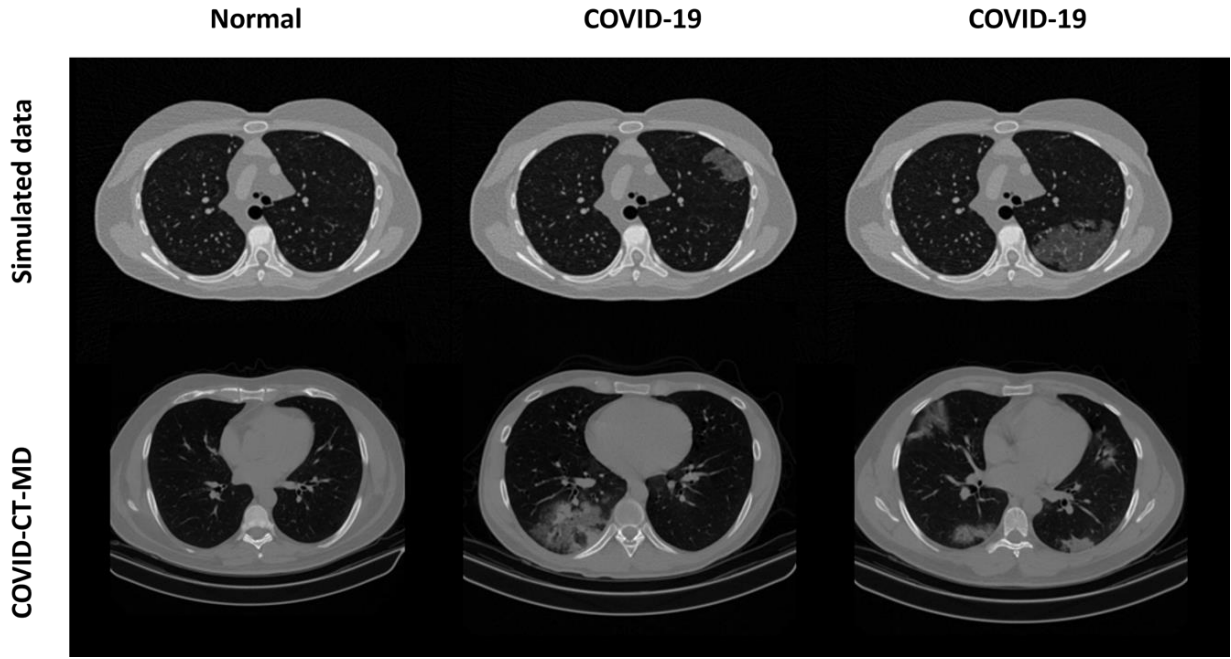

**Figure 2.** Simulated and real CT sample images. The top row shows simulated CVIT-COVID images with three different conditions: normal (left) and two different COVID-19 disease abnormalities (middle and right). The bottom row shows clinical CT images in a similar configuration: normal (left) and two different COVID-19 cases (middle and right) from COVID-CT-MD Dataset [16].

## Deep-learning Model:

This study adopted the proposed deep convolutions neural network (CNN) presented by Gunraj *et al.* to evaluate the performance of virtually imaged simulated CT images [2]. This deep CNN has been designed using a machine-driven design approach. We used the exact implementation made public by the authors at https://github.com/haydengunraj/COVIDNet-CT. The acquired model was pre-trained on the ImageNet [17] dataset and then was trained on the COVIDx-CT [2] dataset. We have analyzed COVID-19 positive vs. negative classification performance. The 2D model performed slice-by-slice prediction. To get a patient-based classification, we averaged the slices with each patient's top 10 percent outputs.

For the in-house model, we have used a lightweight version of the 3D residual convolutional neural network (CNN) introduced in our earlier studies [9, 18]. One initial convolution was performed on input volumes; afterward, features were learned in two resolution scales using single R-block units in each resolution. A R-block consists of batch-normalization, rectified linear unit (ReLu) activation, and 3D convolution [18]. After 2$^{nd}$ resolution last R-block features passed through batch-normalization, ReLu, global average-pooling, dropout (0.5), and finally the sigmoid classification layer for binary prediction (COVID-19 positive vs. negative). Weighted binary cross-entropy loss (W-CEL) was used as a loss function, and the model was trained from scratch using an SGD optimizer with an exponential decay learning rate.

## Results & Discussion:

Overall evaluation was performed at the patient level. Performance was reported as the receiver operating characteristic area under the curve (AUC), with 95% CIs calculated using the DeLong method [19]. We assessed model performance depending on the dataset used for training vs. testing, COVID-19 infection size, and CT dose.

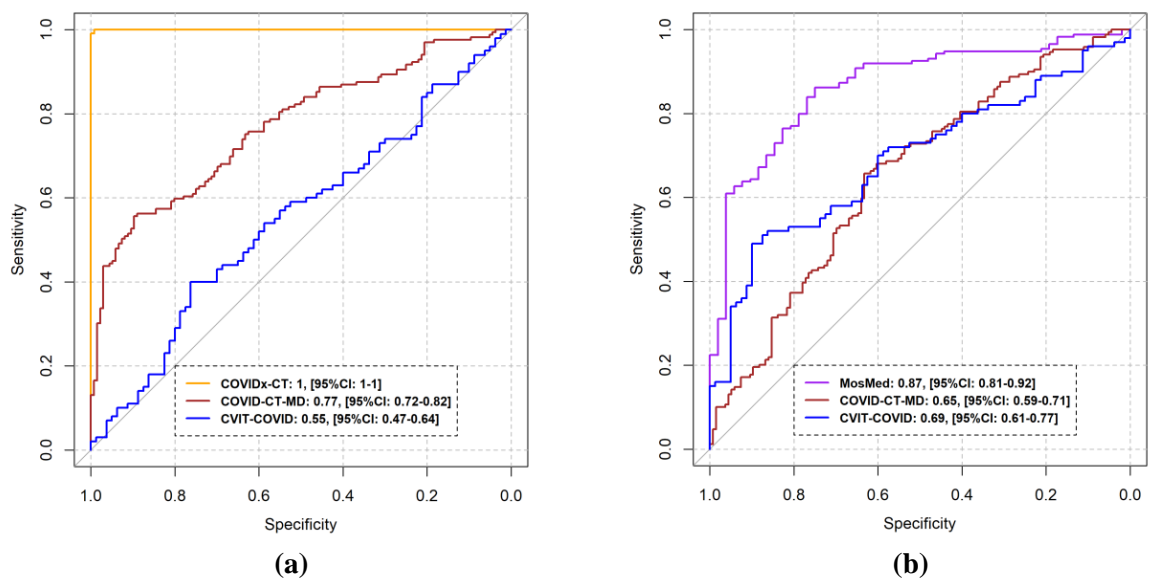

**Figure 3.** Area under the receiver operating characteristic curve (AUC) of internal and external test performance utilizing simulated and real clinical public dataset**. (a)** Patient-level performance of the open-source model on the internal COVIDx-CT test set and external performance on simulated performance on simulated XCAT and clinical COVID-CT-MD dataset. (b) Patient-level performance of the in-house model on the internal MosMed test set and external performance on simulated XCAT and clinical COVID-CT-MD dataset. Note the close correspondence in AUC and curves for the simulated CVIT-COVID (blue) and COVID-CT-MD (brown) datasets.

The open-source model has perfect performance with an AUC of 1 on the internal test set (COVIDx-CT test set), which did not generalize to any of our external tests shown in fig. 3(a). External testing on simulated CVIT-COVID cases yielded 0.55 AUC and 0.77 on the COVID-CT-MD dataset. When our in-house model was evaluated on the internal test set (MosMed test set), the performance was 0.87, which also dropped when tested on simulated CVIT-COVID with AUC of 0.69 and COVID-CT-MD dataset with an AUC of 0.65 (fig. 3(b)). Both models failed to generalize, but the magnitude of the performance drop was less when using our in-house model, and performance was similar in external evaluation of the in-house model on the clinical and simulated dataset. Further analysis was performed utilizing our in-house model.

Our results demonstrated no dose-dependence for the model tested on simulated data shown in fig. 4. We have analyzed performance separately according to the COVID-19 disease extent depending on whether it was greater or lesser than 2.65% of the lungs volume, which is the mode value observed for our simulation cases. Classification performance was substantially higher when a patient had ≥ 2.65% disease extent (AUC=0.87) compared to < 2.65% (AUC=0.52) (fig. 5).

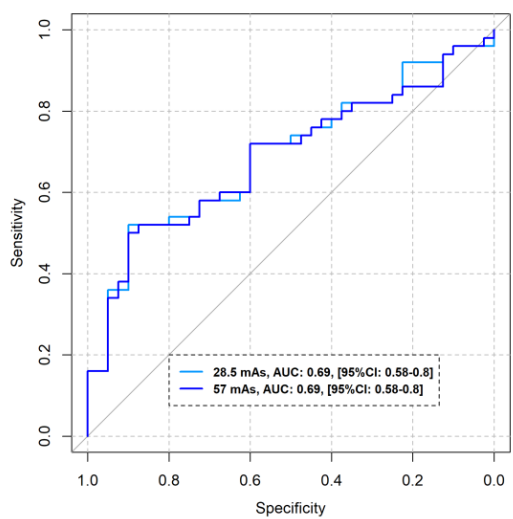

**Figure 4.** Area under the receiver operating characteristic curve (AUC) of in-house model on the simulated CVIT-COVID test dataset under different CT dose levels

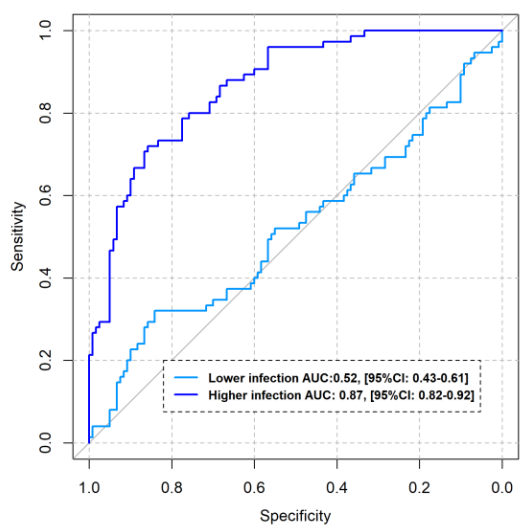

**Figure 5.** Area under the receiver operating characteristic curve (AUC) of in-house model on the simulated CVIT-COVID test dataset under different COVID-19 infection extension based on lungs volume. Note- Lower infection: COVID-19 infection < 2.65% of the lungs volume, Higher infection: COVID-19 infection ≥ 2.65% of the lungs volume.

In this work, we have evaluated an open-source deep-learning classifier trained on multi-institutional data and an in-house model trained on open clinical dataset utilizing the VITs. A notable limitation is that only two models were evaluated, as our initial focus was to demonstrate the utility of our VIT framework, which can rigorously analyze any model in an agnostic and objective way. Although performances dropped considerably relative to the original model's nearly perfect internal test results, those results are almost certainly overtrained. In fact, in a separate study, we showed the same Waterloo group's AI model for classifying COVID-19 for CXR completely failed to generalize, with AUC dropping from perfect to random chance on two different datasets [20]. In addition, the VIT framework allowed comparison of the effects of dose as well as disease extent. Such physical and clinical insights would not have been practicable to achieve using real clinical trials, thus confirming the utility of virtual imaging simulations.

**Acknowledgments:**
This work was funded in part by the Center for Virtual Imaging Trials, NIH/NIBIB P41-EB028744. We thank Lavsen Dahal and Kyle Lafata for helpful discussions.